\documentclass[lettersize,journal]{IEEEtran}
\usepackage{amsmath,amsfonts}
\usepackage{algorithmic}
\usepackage{algorithm}
\usepackage{array}
\usepackage[caption=false,font=normalsize,labelfont=sf,textfont=sf]{subfig}
\usepackage{textcomp}
\usepackage{stfloats}
\usepackage{url}
\usepackage{verbatim}
\usepackage{graphicx}
\usepackage{cite}

\usepackage{makecell}
\usepackage{setspace}
\usepackage{multirow}
\usepackage{diagbox}
\usepackage{bbm}
\usepackage{tcolorbox}
\usepackage{balance}

\newenvironment{mybullet}{\begin{list}{$\bullet$}
		{\setlength{\topsep}{0.5mm}\setlength{\itemsep}{0.5mm}
			\setlength{\parsep}{0.5mm}
			\setlength{\itemindent}{0.2mm}\setlength{\partopsep}{0.5mm}
			\setlength{\labelwidth}{15mm}
			\setlength{\leftmargin}{3mm}}}{\end{list}}

\newcommand{\answer}[2]
{
	\begin{tcolorbox}[boxrule=1pt,left=2pt, right=2pt, top=2pt,bottom=2pt,]
		\textbf{Answer to RQ#1:} #2
	\end{tcolorbox}
}

\begin{document}

\title{$\mathbb{ABC-}$$\mathbbm{Channel}$: An Advanced Blockchain-based Covert Channel}

\author{\IEEEauthorblockN{
        Xiaobo Ma\IEEEauthorrefmark{1}\IEEEauthorrefmark{2}, 
        Pengyu Pan\IEEEauthorrefmark{1}\IEEEauthorrefmark{2},
        Jianfeng Li\IEEEauthorrefmark{1}\IEEEauthorrefmark{2}, 
        Wei Wang\IEEEauthorrefmark{3},
        Weizhi Meng\IEEEauthorrefmark{4},
        Xiaohong Guan\IEEEauthorrefmark{1}\IEEEauthorrefmark{2}
        \\
	\IEEEauthorblockA{\IEEEauthorrefmark{1}MOE Key Lab for Intelligent Networks and Network Security, Xi'an Jiaotong University,  Xi'an, China}\\
	\IEEEauthorblockA{\IEEEauthorrefmark{2}Faculty of Electronic and Information Engineering, Xi'an Jiaotong University, Xi'an, China}\\
    \IEEEauthorblockA{\IEEEauthorrefmark{3}Beijing Jiaotong University, Beijing, China}\\
    \IEEEauthorblockA{\IEEEauthorrefmark{4}Technical University of Denmark, Kongens Lyngby, Denmark}\\
	}	
	}



\maketitle

\begin{abstract}
 Establishing efficient and robust covert channels is crucial for secure communication within insecure network environments. 
 With its inherent benefits of decentralization and anonymization, blockchain has gained considerable attention in developing covert channels.
 To guarantee a highly secure covert channel, channel negotiation should be contactless \textit{before} the communication, carrier transaction features must be indistinguishable from normal transactions \textit{during} the communication, and communication identities must be untraceable \textit{after} the communication.
 Such a full-lifecycle covert channel is indispensable to
 defend against a versatile adversary who intercepts two communicating parties comprehensively (e.g., on-chain and off-chain). 
 Unfortunately, 
it has not been thoroughly investigated in the literature.
We make the first effort to achieve a full-lifecycle covert channel, a novel blockchain-based covert channel named \texttt{ABC-Channel}. 
We tackle a series of challenges,
such as off-chain contact dependency,  
increased   
masquerading difficulties as growing transaction volume,
and time-evolving, communicable yet untraceable  identities,
to achieve contactless channel negotiation,  indistinguishable transaction features, and untraceable communication identities, respectively.
 We develop a working prototype to validate \texttt{ABC-Channel} and conduct extensive tests on the Bitcoin testnet. 
The experimental results demonstrate that  \texttt{ABC-Channel} achieves substantially secure covert capabilities. 
In comparison to existing methods, it also exhibits state-of-the-art transmission efficiency.
 
\end{abstract}


\section{Introduction}
\label{Introduction}
 
 Covert channels have long been a focal point of research due to their utility in secure message transmission within insecure network environments. 
 Traditionally, covert channels are categorized into covert storage channels (CSC) and covert timing channels (CTC) based on the message carrier  \cite{tian2020survey}. 
 CSC utilizes storage fields as message carriers \cite{mileva2021comprehensive, binsalleeh2014characterization, nazari2020lightweight}.
 It is exemplified by the widely known Tor network, acclaimed for its anonymity. 
 However, extensive research reveals that  CSC faces significant threats, including fingerprinting attacks and potential deanonymization \cite{zhan2022detecting, nasr2018deepcorr, karunanayake2021anonymisation, oh2022deepcoffea}.
 CTC, on the other hand, employs time series as message carriers \cite{cabuk2004ip, seong2022practical, zhu2023novel}. 
 However, this approach is highly susceptible to network fluctuations and noises, resulting in insufficient robustness \cite{cao2020chain}.
 Therefore, addressing these limitations necessitates the development of innovative covert channel solutions.
 
 Since the introduction of Bitcoin by Satoshi Nakamoto in 2009,   blockchain technology has garnered significant attention  \cite{bitcoin}.
 Due to its inherent advantages of decentralization and anonymity, 
 blockchain has been extensively studied and utilized for various purposes, including serving as command and control (C\&C) channels for botnets and enabling covert communication \cite{zhang2023covert, ali2018zombiecoin, frkat2018chainchannels, chen2022blockchain, zheng2021keys}.
 Public chains like Bitcoin, which are globally accessible and operate without censorship, allow users to send transactions anytime and anywhere without being audited. 
 Therefore, using blockchain as a carrier for covert channels presents inherent advantages.
 
Despite the inherent advantages,
implementing a highly secure   blockchain-based covert channel  is challenging 
because of 
the following three essential requirements.


\noindent
\textbf{R1: Contactless Channel Negotiation.}
Before establishing a blockchain-based covert channel, the communicating parties must reach a consensus on channel negotiation information (e.g., transaction addresses, symmetric keys). 
 To securely transmit this information, the two parties need to contact each other through an alternative channel (e.g.,  end-to-end secure transmission, third-party relay), namely,  an off-chain channel.
 We refer to such off-chain communication over the network as contact channel negotiation.
 While contact channel negotiation ensures information confidentiality, it does not guarantee anonymity.
 In an insecure network, the communication parties could be exposed, even when using anonymization techniques like Tor, which may suffer from de-anonymization and tracing attempts \cite{karunanayake2021anonymisation}. 
 Therefore, it becomes imperative to establish a covert channel through contactless channel negotiation, which minimizes exposure risks.

\noindent
\textbf{R2: Indistinguishable Transaction Feature.} 
 Utilizing blockchain-based covert channels for high-throughput covert communication entails generating numerous carrier transactions.
 Manually constructing the parameters of these transactions, including input amount and transaction fee, can introduce common features among them. 
  Since blockchain-based covert channels employ transactions as message carriers, such common features may expose the channel to attackers. 
 Therefore, it is necessary to generate carrier transactions with indistinguishable features from normal transactions.

\noindent
\textbf{R3: Untraceable Communication Identity.} 
 While the blockchain provides a degree of anonymity, it remains to trace the identities of address holders through the analysis of multiple transactions associated with the same address \cite{toyoda2019novel,ermilov2017automatic,xiang2022babd}. 
 Users' repetitive use of the same address increases the exposure of address-related features, such as average spending and transaction sending cycles. 
 As all transactions are permanently stored on the blockchain, attackers can access and analyze the complete historical transactions. If the sender's address is detected, the covert channel is exposed. 
 Therefore, it is crucial for blockchain-based covert channels to employ dynamic transaction addresses, ensuring that communication identities cannot be traced through fixed addresses.

 To achieve the above three key requirements,
 we propose \texttt{ABC-Channel}, a novel covert channel that facilitates high-throughput covert communication within insecure network environments.

 Achieving \textbf{R1} is not easy.
 The major challenge is off-chain contact dependency. 
  Without such off-chain contact,  the two parties cannot 
 establish the covert channel further. 
 To elude the need for off-chain contact, we develop a contactless sharing method. 
  This method allows the two parties to generate the same negotiation information and subsequently establish the covert channel without any contact, thereby making all communication conducted purely based on blockchain.
The only prerequisite of this method for the two parties is kept at a minimum, i.e.,   knowing each other's public keys, which are publicly available on the blockchain. 


  Meeting  \textbf{R2} needs to figure out a 
  transaction feature masquerading mechanism that 
  ensures indistinguishability between carrier transactions and normal transactions.
  To achieve this objective, existing research uses the nonce in digital signatures as a carrier for covert messages.
  However, 
 only using the nonce as carrier has 
  a realistic obstacle. 
That is,  
as transaction volume increases, it becomes more difficult to conceal common features like transaction fee and the number of inputs among carrier transactions, leading to increased transaction masquerading difficulties.
To obfuscate these features,
we propose a method based on GAN to disguise the parameters of the carrier transactions so that   carrier transactions and normal transactions are indistinguishable at a large scale. 

 
 To implement \textbf{R3},
 a straightforward approach is to use unique transaction addresses for each carrier transaction. 
 This approach calls for 
synchronizing each carrier transaction's addresses between the two parties.
We design a dynamic address synchronization mechanism that 
 enables time-evolving,
communicable, yet untraceable identities.
The mechanism allows the sender to constantly changing the sending/receiving address of each carrier transaction. Note that, in our design, the receiving address is randomly generated, and the receiver only needs to synchronize the sending address to be aware that the carrier transaction is for him/her, making no one except the receiver knows this transaction is for him/her. Although the receiver does not have the private key of the carrier transaction with randomly generated receiving address,  he/she can still decrypt the  carrier transaction due to our hierarchal deterministic-based private key derivation design.

 \begin{table}[htp]
 \vspace{-7pt}
 \caption{Comparison of \texttt{ABC-Channel} and other methods}
 \vspace{-5pt}
 \label{tab:ABC-Channel vs other methods}
 \centering
 \small
 \setlength{\tabcolsep}{15pt}
 \begin{tabular}{lccc}
 \hline
 Literature & \textbf{R1} & \textbf{R2} & \textbf{R3}  \\ \hline
 
 \makecell[l]{Gao et al. \cite{gao2020achieving}} & $\times$ & $\times$ & \checkmark \\ 
 
 \makecell[l]{MRCC \cite{guo2021mrcc}} & $\times$ & $\times$ & \checkmark \\ 
 
 \makecell[l]{DLchain \cite{tian2020dlchain}} & $\times$ & $\times$ & \checkmark \\

 \makecell[l]{BLOCCE \cite{partala2018provably}} & $\times$ & $\times$ & $\times$ \\
 
 \makecell[l]{Cao et al. \cite{cao2020chain}} & $\times$ & $\times$ & \checkmark \\
 
 \makecell[l]{MSCCS \cite{liu2022msccs}} & $\times$ & $\times$ & \checkmark \\ 

 \makecell[l]{Tiemann et al. \cite{tiemann2021act}} & \checkmark & $\times$ & \checkmark \\ \hline
 
 \makecell[l]{\texttt{ABC-Channel}} & \checkmark & \checkmark & \checkmark \\ \hline

 \end{tabular}
 \end{table}

 
 
 

 
 

 


 Our primary contributions can be summarized as follows.
 \begin{mybullet}
     \item 
     To our knowledge, \texttt{ABC-Channel} is the first blockchain-based covert channel that fulfills the contactless,  indistinguishable, and untraceable requirements while enabling high-throughput communication.
     In Table \ref{tab:ABC-Channel vs other methods}, we compare \texttt{ABC-Channel} with existing methods.
     
     \item 
     To achieve the three requirements, we  propose  three solutions:   contactless sharing, GAN-based transaction feature
masquerading, and dynamic
address synchronization. 
     These solutions enable the sender and receiver to communicate covertly and establish dynamic address synchronization without direct contact.
    
     \item 
     We implemented a prototype of \texttt{ABC-Channel} on the Bitcoin testnet and evaluated its effectiveness. 
     Our evaluation assumes a strong attacker that knows our design details. 
    The precision and recall for the attack in identifying \texttt{ABC-Channel}  are only 0.549 and 0.491, respectively, which is approximately equivalent to random guessing.
    Each transaction with $n$ inputs in \texttt{ABC-Channel} can carry a $256\times n$-bit message, a SOTA level compared to other methods using implicit embedding in Bitcoin. 


     

 \end{mybullet}
 
 \noindent
 \textbf{Roadmap.}
 Sec. \ref{Background And Threat Model} introduces  the background  and    threat model.
 In Sec. \ref{System Design}, we   detail the  design of \texttt{ABC-Channel} and Sec. \ref{Evaluation}  implements, verifies, and evaluates  it on the Bitcoin testnet.
 Sec. \ref{Discussion} discusses potential challenges.
 Sec. \ref{Related Work} surveys the literature, and we finally conclude in Sec. \ref{Conclusion}.
\section{Background And Threat Model}
\label{Background And Threat Model}
 
\subsection{Background}
\label{Background}
 
\noindent
$\bullet$
\textbf{HD Wallet.}
The Hierarchical Deterministic (HD) wallet is  for efficient transaction address management and can be implemented in many blockchains \cite{bip32, bip44}.
 It works as follows.
 First, the user generates a random number as a seed. 
 Then, the HMAC-SHA512 algorithm is employed to compute the hash value of the seed. 
 The leftmost 256 bits of the resulting hash are utilized as the private key ($sk$), while the rightmost bit is used as the chain code ($chaincode$). 
 The $chaincode$ adds randomness to the derivation of keys and can be replaced by users if desired. 
 The combination of $sk$ and $chaincode$ is an extended private key, denoted as $Esk$. 
 By utilizing $Esk$ and an index, users can derive both the child $sk$ and the child $chaincode$.
 HD wallet will be used to create a dynamic communication identity.


\noindent
$\bullet$
\textbf{Kleptography.}
 This is an attack on asymmetric cryptography, proposed by Young and Yung \cite{young1997kleptography}. 
 A two-step signature process allows a sender to clandestinely transmit his private key ($sk_{sender}$) to the receiver.
 The sender first signs data 1 to obtain signature 1. 
 Then, using kleptography, the sender signs data 2, incorporating the nonce from signature 1 and the receiver's public key ($pk_{receiver}$) as additional inputs.
 To recover $sk_{sender}$, the receiver uses signature 1, signature 2, data 2, and his private key ($sk_{receiver}$) as inputs.
 This covert mechanism exploits vulnerabilities in asymmetric cryptography (e.g., ECDSA), enabling surreptitious transmission of private keys between sender and receiver.
  Inspired by this approach, we devise a contactless channel negotiation method for blockchain-based covert channels.

\noindent
$\bullet$
\textbf{Subliminal Channel.}
 Subliminal channels serve as a means to transmit messages securely within insecure network environments. 
 In 1984, Gustavus J. Simmons \cite{simmons1984prisoners} proposed a subliminal channel specifically designed for digital signatures.
 In this method, the sender replaces the nonce within the signature with the intended message and signs it using his private key ($sk_{sender}$). 
 Consequently, when the receiver possesses both $sk_{sender}$ and the signature, he can extract the message from the sender. 
 Using this method in the blockchain, we can avoid modifying the transaction fields.
 Inspired by this, we implement indistinguishable transaction features when performing high-throughput covert communication.
 

\noindent
$\bullet$
\textbf{ECDH.}
 The Elliptic Curve Diffie–Hellman (ECDH) algorithm is devised to facilitate the generation of a shared key ($Key_{share}$) among communicating parties within an insecure network environment. 
 From the perspective of the sender, this computational process can be succinctly represented as $Key_{share} = ECDH(sk_{sender}, pk_{receiver})$. 
 Here, $sk_{sender}$ denotes the sender's private key, while $pk_{receiver}$ represents the receiver's public key. 
 Similarly, using an analogous procedure, the receiver can compute $Key_{share}$. 
 In this study, we employ the ECDH algorithm with SHA256 to derive the chaincode, expressed as $chaincode = SHA256(Key_{share})$.


\subsection{Threat Model}
\label{Threat Model}
\begin{figure}[htp]
 \centering 
 \includegraphics[width=1.0\columnwidth]{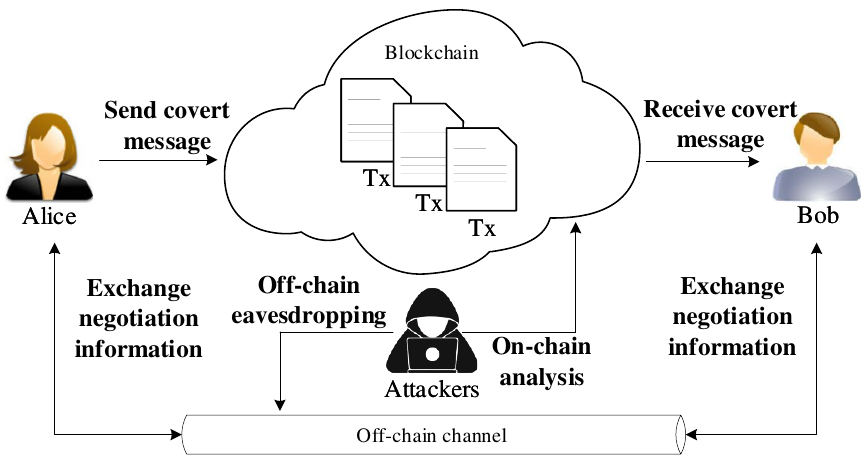}
 \vspace{-0.8cm}
 \caption{\label{fig:threat model}Blockchain-based covert channel threat model.}
 \end{figure}

  Fig. \ref{fig:threat model} illustrates the threats encountered by blockchain-based covert channels, wherein Alice symbolizes the sender, and Bob signifies the receiver. 
  The predominant threats encompass adversaries monitoring off-chain channels and scrutinizing transactions. 
  To establish a clear framework, we define specific assumptions and limitations regarding the attackers' capabilities as follows.

 \noindent
 $\bullet$
 \textbf{Off-chain Eavesdropping.}
 In an insecure network environment, attackers can eavesdrop on the communication between Alice and Bob but cannot execute a man-in-the-middle attack.
 They possess the capability to detect any form of contact communication and can utilize diverse methods such as fingerprinting.
 Even in scenarios where attackers cannot decrypt the ciphertext, they retain the ability to ascertain the identities of the communicating parties.
 
 
 \noindent
 $\bullet$
 \textbf{On-chain Analysis.}
 Attackers are also capable of scrutinizing transactions on the blockchain, which involves thoroughly examining transactional data and tracking addresses. 
 They can investigate various transaction attributes, including the transaction amount, timestamp, and the usage of fields, to identify transactions that deviate from normal patterns. 
 Furthermore, when an address is linked to a significant number of transactions, attackers can infer its critical characteristics. 
 By analyzing these characteristics, attackers can infer the type of address and might even trace the identity of its owner.
 

\section{System Design}
\label{System Design}
 This section details the design of \texttt{ABC-Channel}.
 Table \ref{tab:major notation} defines major notations.

 \begin{table}[htp]
 \footnotesize
 \normalsize
 \centering
 \vspace{-7pt}
 \caption{Major notations}
 \vspace{-5pt}
 \label{tab:major notation}
 \resizebox{\columnwidth}{!}{
 \begin{tabular}{l|l}
 \hline
 \textbf{Notation} & \textbf{Definition}\\ \hline
 $Tx_k$ & Transaction set signed using kleptography\\
 
 $Tx_k^i$ & \makecell[l]{The $i$th transaction of $Tx_k$}\\
 
 $Tx_c$ & Transaction set signed using subliminal channel\\ 
 
 $Tx_c^i$ & The $i$th transaction of $Tx_c$\\

 $(pk_{Alice}, sk_{Alice})$ & Alice's public and private key pair \\
 
 $(pk_{Bob}, sk_{Bob})$ & Bob's public and private key pair \\
 
 $(pk_c^i, sk_c^i)$ & The key pair used for sending $Tx_c^i$\\
 
 $addr_c^i$ & The address corresponding to $sk_c^i$\\
 
 $Esk_{AB}$ & Extended sk based on $sk_{Alice}$ \\
 \hline
 
 \end{tabular}
 }
 \vspace{-10pt}
 \end{table}

\subsection{System Overview}
\label{System Overview}

\begin{figure*}[htp]
 \centering 
 \includegraphics[width=1.95\columnwidth]{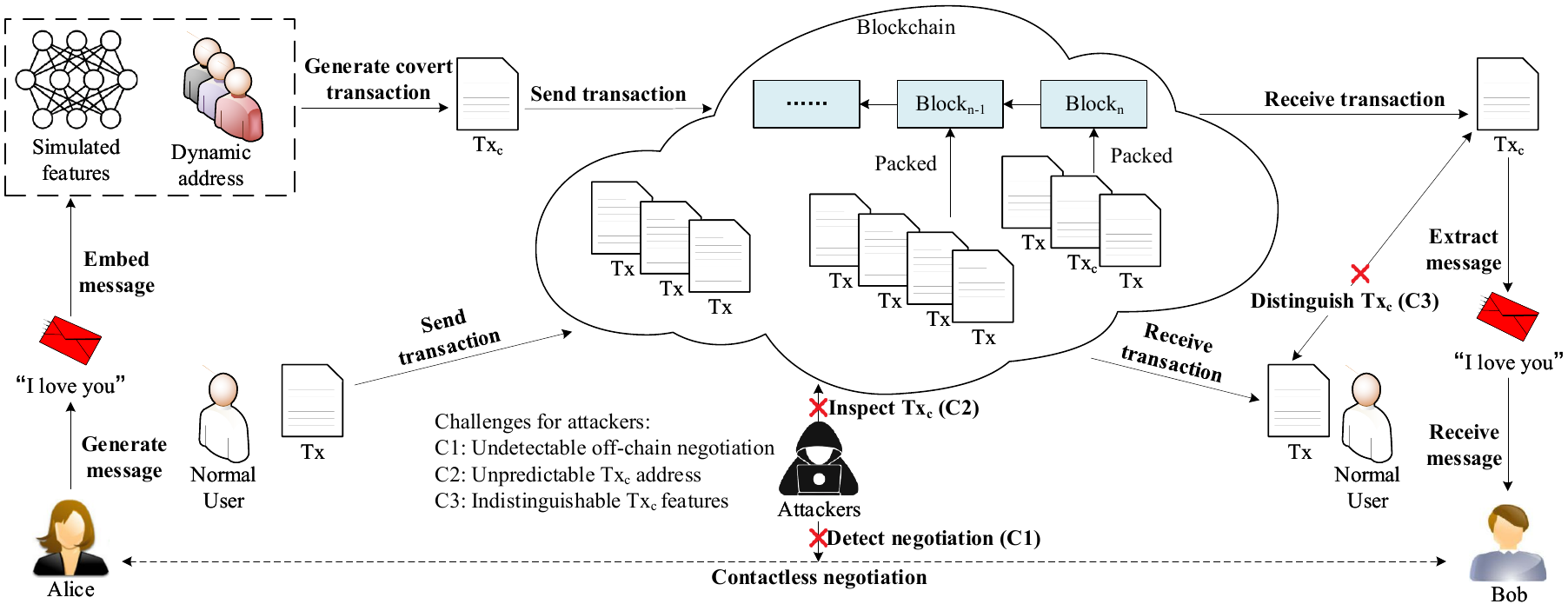}
 \vspace{-0.3cm}
 \caption{\label{fig:architecture}\texttt{ABC-Channel} architecture.}
 \vspace{-12pt}
 \end{figure*}

 In the scenario where communication security cannot be guaranteed, Alice and Bob can utilize \texttt{ABC-Channel} for covert communication.
 It is presupposed that they are privy to each other's public keys.
 To delineate ownership, we employ subscripts to denote each party's public key, private key, and address.
 In this paper, we classify transactions carrying information as carrier transactions. 
 These carrier transactions are divided into key transactions ($Tx_k$) for transmitting negotiation information used in building the covert channel and covert transactions ($Tx_c$) for conveying covert messages.
 All other transactions are considered normal transactions.
 The workflow of \texttt{ABC-Channel}, as depicted in Fig. \ref{fig:architecture}, encompasses three main components: on-chain covert communication, simulated transaction features, and dynamic address generation.

\noindent
$\bullet$
\textbf{On-chain Covert Communication.}
 In the \texttt{ABC-Channel}, 
 Alice employs $Tx_c$ for message transmission. 
 The procedure initiates when Alice inputs the message into \texttt{ABC-Channel}, triggering the generation of a covert transaction, referred to as $Tx_c$.
 \texttt{ABC-Channel} generates simulated transaction characteristics along with an address specific to $Tx_c$. 
 Subsequently, $Tx_c$ is dispatched to the blockchain. 
 Upon reception, Bob can extract the message from Alice. 
 Significantly, only Bob can identify $Tx_c$ as distinct, with attackers being unable to differentiate it from normal transactions on the blockchain. 
 
\noindent
$\bullet$
\textbf{Simulated Transaction Features.} 
 To augment the resemblance of $Tx_c$ to normal transactions, \texttt{ABC-Channel} utilizes simulated features for $Tx_c$ derived from the characteristics of standard normal transactions. 
 Leveraging historical transactions as a training dataset, \texttt{ABC-Channel} learns and integrates features such as transaction fees and input amounts. 
 This approach enables $Tx_c$ to mimic the attributes of normal transactions closely. 
 Sec.  \ref{Indistinguishable Transaction Feature} presents a detailed exploration of this methodology.
 We will provide a detailed evaluation of this method's effectiveness in Sec. \ref{Evaluation}.

\noindent
$\bullet$
\textbf{Dynamic Address Generation.}
 To enhance the untraceability of \texttt{ABC-Channel}, dynamic address generation is employed instead of using fixed addresses. 
 Each $Tx_c$ will utilize a unique address as the sending address. 
 The focus is primarily on the sending address, as the receiving address does not impact the functionality of \texttt{ABC-Channel}.
 Alice can generate additional receiving addresses, as long as she ensures that each receiving address is unique.
 By dynamically generating addresses for each $Tx_c$, attackers cannot associate multiple $Tx_c$ through a single address. 
 Meanwhile, the address used by the $Tx_c$ is unpredictable.
 Sec. \ref{Untraceable Communication Identity} will discuss this method in detail.

 This subsection outlines how Alice sends messages to Bob using \texttt{ABC-Channel}. 
 They synchronize negotiation information through contactless negotiation, specifically using $Tx_k$ as explained in Sec. \ref{Contactless Channel Negotiation}. 
 $Tx_k$ consists of only two transactions, while the number of $Tx_c$ depends on the number of messages transmitted.  
 Alice's address is only in $Tx_k$, while Bob's address does not appear in $Tx_k$ and $Tx_c$.
Note that $Tx_c$ uses the derived address as the sending address.
 
 

 \subsection{Contactless Channel Negotiation}
 \label{Contactless Channel Negotiation}
 
 To address \textbf{R1}, we propose a kleptography-based contactless channel negotiation approach. 
 In previous blockchain-based covert channel approaches, it is common to utilize off-chain communication to synchronize negotiation information before establishing covert channels. 
 However, previous methods often involved direct communication between the parties, relying on encryption to prevent information leakage. 
 Unfortunately, this direct communication approach can inadvertently expose the existence of communication and potentially reveal the real identities of the involved parties. 
 The objective of this subsection is to enable covert completion of negotiation information  (i.e., $Esk_{AB}$) synchronization between the two parties without the need for direct communication.

\noindent
$\bullet$
\textbf{Private Key Embedding.}
 The process for generating the extended private key $Esk_{AB}$ entails the use of Alice's private key ($sk_{Alice}$) and the $chaincode_{AB}$. 
 The private key $sk_{Alice}$ is conveyed to Bob through the transaction $Tx_k$, as outlined in Algorithm \ref{alg:privkey embeded}. 
 To facilitate understanding, we introduce two hypothetical participants, Charlie and Dave. 
 In this context, Alice uses kleptography to construct transaction $Tx_k$, assigning $addr_{Charlie}$ and $addr_{Dave}$ as the recipients of $Tx_k$. It is imperative to emphasize that Charlie and Dave do not participate in the covert communication between Alice and Bob and are unaware of the exchange.
 

 Alice generates the raw transaction bytes, denoted as $RawTx_k^1$ and $RawTx_k^2$, using the $GenRawTx$ function. 
 For $RawTx_k^1$, she employs the standard signature algorithm ($Sign$) to sign it while utilizing the $GetNonce$ function to acquire the random number ($nonce_k^1$) used during the signing process.  
 For $RawTx_k^2$, Alice utilizes the $SignByKlepto$ function, which incorporates kleptography. 
 Compared to $Sign$, $SignByKlepto$ requires additional inputs, namely $nonce_k^1$ and $pk_{Bob}$.
 Using the generated transaction raw bytes and digital signatures, Alice employs the $GenTx$ function to generate $Tx_k^1$ and $Tx_k^2$.
 \begin{algorithm}
 \footnotesize
 \setstretch{1.35}
 \renewcommand{\algorithmicrequire}{\textbf{Input:}}
 \renewcommand{\algorithmicensure}{\textbf{Output:}}
 \caption{Embed the private key into $Tx_k$} 
 \label{alg:privkey embeded} 
     \begin{algorithmic}[1]
        \REQUIRE $sk_{Alice}$, $pk_{Bob}$, $addr_{Charlie}$, $addr_{Dave}$
    	\ENSURE $(Tx^1_k, Tx^2_k)$
     
        \STATE $addr_{Alice} \gets GenAddrFromSk(sk_{Alice})$
        
        \STATE $RawTx_k^1 \gets GenRawTx(addr_{Alice}, addr_{Charlie})$
        \STATE $RawTx_k^2 \gets GenRawTx(addr_{Alice}, addr_{Dave})$
        
    	\STATE $(r, s)_k^1 \gets Sign(sk_{Alice}, RawTx_k^1)$ 
        
        \STATE $nonce_k^1 \gets GetNonce((r, s)_k^1)$    	
        \STATE $(r, s)^2_k \gets SignByKlepto(sk_{Alice}, pk_{Bob}, nonce_k^1,  RawTx_2)$ 

        \STATE $Tx_k^1 \gets GenTx(RawTx_k^1, (r, s)_k^1)$ 
    	\STATE $Tx_k^2 \gets GenTx(RawTx_k^2, (r, s)_k^2)$ 
            
        \RETURN $(Tx^1_k, Tx^2_k)$
    		
    \end{algorithmic} 
\end{algorithm}

\noindent
$\bullet$
\textbf{Private Key Extraction.}
 As described in Algorithm \ref{alg:privkey extract}, Bob retrieves $sk_{Alice}$ from $Tx_k$ using the following steps.
 First, Bob searches for the transaction $Tx_k$ on the blockchain using the function $GetTxFromAddr$.
 Then, the function $GetSignFromTx$ is applied to obtain the digital signatures $(r,s)_k^1$ and $(r,s)_k^2$ from $Tx_k$.
 Additionally, Bob utilizes the function $GetRawTx$ to retrieve the raw byte representation of $Tx_k^2$, denoted as $RawTx_k^2$.
 Finally, Bob provides $sk_{Bob}$, $pk_{Bob}$, and the information above as input to the function $GetSk$ to extract $sk_{Alice}$.

 \begin{algorithm} 
 \footnotesize
 \setstretch{1.35}
 \renewcommand{\algorithmicrequire}{\textbf{Input:}}
 \renewcommand{\algorithmicensure}{\textbf{Output:}}
 \caption{Extract the private key from $Tx_k$} 
 \label{alg:privkey extract} 
     \begin{algorithmic}[1]
         \REQUIRE $sk_{Bob}$, $pk_{Bob}$, $pk_{Alice}$
         \ENSURE $sk_{Alice}$

         \STATE $addr_{Alice} \gets GenAddrFromPk(pk_{Alice})$
         \STATE $(Tx_k^1, Tx_k^2) \gets GetTxFromAddr(addr_{Alice})$
         \STATE $(r, s)_k^1 \gets GetSignFromTx(Tx_k^1)$
         \STATE $(r, s)_k^2 \gets GetSignFromTx(Tx_k^2)$

         \STATE $RawTx_k^2 \gets GenRawTx(Tx_k^2)$
         
         \STATE $sk_{Alice} \gets GetSk(sk_{Bob}, pk_{Bob}, (r, s)_k^1, (r, s)_k^2, RawTx_k^2)$ 
           
         \RETURN $sk_{Alice}$
     \end{algorithmic} 
 \end{algorithm}

\noindent
$\bullet$
\textbf{Negotiation Information Generation.}
 Both Alice and Bob utilize the Elliptic-curve Diffie–Hellman (ECDH) algorithm and SHA256 hashing function to generate $chaincode_{AB}$. 
 By utilizing ($sk_{Alice}$, $pk_{Bob}$) and ($sk_{Bob}$, $pk_{Alice}$) respectively in the ECDH computation, they each derive the same shared key. 
 Subsequently, applying the SHA256 hash function to the shared key yields $chaincode_{AB}$. 
 With the consistent utilization of $sk_{Alice}$ and $chaincode_{AB}$, both Alice and Bob can concurrently generate the same extended private key $Esk_{AB}$.
 Attackers are unable to obtain $Esk_{AB}$ since they lack access to $sk_{Alice}$ and $chaincode_{AB}$.

\subsection{Indistinguishable Transaction Feature}
\label{Indistinguishable Transaction Feature}
\begin{figure}[htp]
 \centering 
 \includegraphics[width=0.8\columnwidth]{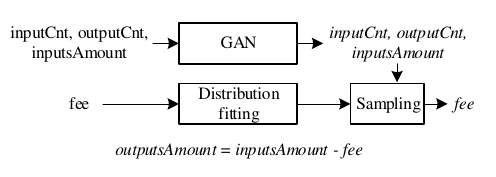}
 \vspace{-5pt}
 \caption{\label{fig:feature generation}Parameter  generation (\textit{Italics}:   generated using our method).}
 \vspace{-10pt}
 \end{figure}

 To address \textbf{R2}, we propose a mixed strategy-based approach for generating transaction parameters, using Bitcoin as an illustrative example.
 Our method focuses on generating five key parameters for covert transactions.
 The $m$th transaction can be expressed by 
 \begin{equation}
 \label{txm}
 \footnotesize
 \begin{split}
 T_m = (&inputCnt_m, outputCnt_m, fee_m, \\&inputsAmount_m, outputsAmount_m).
 \end{split}
 \end{equation}
 InputCnt and outputCnt represent a transaction's input and output quantities, respectively.
 InputsAmount and outputsAmount indicate the total amounts involved in the input and output of a transaction, respectively.
 Fee refers to the payment made to miners for transaction processing.
 In Sec. \ref{More transaction Fields}, we explain why these five parameters are the primary focus of our approach.
 As shown in Fig. \ref{fig:feature generation}, we show how to generate the covert transaction parameters using the real transaction parameters.
 Our approach consists of three main components. 

\noindent
$\bullet$
\textbf{GAN Network Training.}
 Firstly, we leverage CTGAN, a specialized GAN network designed for generating tabular data \cite{SDV}, to generate three parameters: inputCnt, outputCnt, and inputsAmount. 
 To train the model, we utilize real transactions (i.e., real data) as the training dataset, thereby ensuring the generation of realistic transaction parameters. 
 The trained model enables the generation of covert transactions with solely these three parameters, which we refer to as \textit{fake data}.

\noindent
$\bullet$
\textbf{Discrete Distribution Fitting.}
 Next, we perform a fitting analysis on the distribution of transaction fees within real data.
 To achieve this, We use the set $K$ to denote all real data.
 We group transactions based on the inputCnt and outputCnt parameters, denoted as $K_{ij}$ for transactions with $i$ inputs and $j$ outputs. 
 \begin{equation}
 \label{kij}
 \footnotesize
 K_{ij} = \{T_m \in K : (inputCnt_m = i \land outputCnt_m = j) \}.
 \end{equation}
 We further divide each transaction set into n intervals based on the percentile of inputsAmount. 
 We denote the collection of inputsAmount in the $n$th interval as $A^n$, with $A^n_{min}$ and $A^n_{max}$ representing the minimum and maximum values within this collection, respectively.
 The $n$th interval in $K_{ij}$ can be represented as 
 \begin{equation}
 \label{kijn}
 \footnotesize
 \begin{aligned}
 K_{ij}^n = \{ T_m \in K_{ij} : A^n_{min} < inputsAmount_m \leq A^n_{max} \}.
 \end{aligned}
 \end{equation}
 The transaction fees within \(K^n_{ij}\) can be represented as
\begin{equation}
 \label{nijn}
 \footnotesize
 \begin{split}
 F_{ij}^n & = \{ fee_m \in T_m : T_m \in K_{ij}^n \}. 
 \end{split}
 \end{equation}
 We perform a discrete distribution fitting on $F^n_{ij}$ and generate $Len^n_{ij}$ transaction fees from it, denoted as $X^n_{ij}$. 
 Here, $Len^n_{ij}$ is used to represent the number of transactions within $K^n_{ij}$.

\noindent
$\bullet$
\textbf{Transaction Feature Sampling.}
 Finally, we associate the three parameters of each covert transaction in the fake data with the corresponding $X_{ij}^n$. 
 Random selection is employed to assign a fee value to each transaction. 
 It is worth noting that in a transaction, the outputsAmount is obtained by subtracting the fee from the inputsAmount. 
 With this step, the generation of the five parameters in the fake data is complete. 
 To optimize the accuracy and realism of our fake data, our methodology extends beyond the mere use of GAN or distribution fitting. 
 Instead of directly sampling real transaction parameters from the dataset, we adopt a more nuanced approach.
 Next, we elaborate on the rationale behind this approach.

 Taking datasets BTD129 and BTD30 as examples (see Sec. \ref{Data Collection} for details), it is observed that inputsAmount and fee exhibit a wide range of values, with 99th percentiles of 15,666,023,540 and 31,671, respectively.
 Considering the sparse distribution resulting from the extensive range of inputsAmount in the datasets containing approximately 6.8 million transactions, relying solely on the GAN to generate all parameters does not yield satisfactory results. 
 The reason is that GAN struggles to effectively learn the relationships between various parameters.
 Similarly, we have not performed distribution fitting for each parameter individually.
 Directly sampling parameters from real data would result in many covert transactions having the same parameters as real transactions, drawing the attackers' attention.
 In contrast to inputsAmount, the distribution of fee is not considered sparse. 
 This implies that many transactions will have identical transaction fees. 
 Therefore, we only fit and sample from the distribution of fee, aiming to achieve more realistic simulation of transaction parameters, while simultaneously reducing the attention of attackers.

 \subsection{Untraceable Communication Identity}
 \label{Untraceable Communication Identity}

 To address \textbf{R3}, Alice and Bob can utilize dynamic identities, enabling them to dynamically generate a private key for each covert transaction. 
 Utilizing the same $Esk_{AB}$, they can produce identical child private keys using the same index.
 In \texttt{ABC-Channel}, the index begins at 0 and increments by one for each covert transaction.
 In blockchains, private keys can be used to derive public keys and addresses.
 The transaction address consists of both the sending and receiving addresses.
 Here, the term transaction address encompasses both the sending and receiving addresses, particularly emphasizing the \textit{sending address} when referring to the \textit{address}.

\noindent
$\bullet$
\textbf{Covert Message Embedding.}
 We demonstrate how to utilize covert transactions to transmit messages in the context of a transaction with only one input.
 Algorithm \ref{alg:message embedding} outlines the process by which Alice embeds messages into covert transactions. 
During the initialization phase, $index_{last}$ denotes the index after the prior communication. 
 $Index_{last}$ is set to 0 for the first communication. 
 To begin, Alice divides the message into segments using $MsgProcessing$, resulting in the set $msg=\{msg_1, msg_2,..., msg_n\}$, each segment being 32 bytes. 
 Subsequently, Alice generates the private key $sk_c^{index}$ using $GenKeyByIndex$. 
 Using $sk_c^{index}$ and $GenTxBySubliminal$, Alice embeds each $msg_i$ into $Tx_c^{index}$. 
 This process culminates in creating the covert transaction set $Tx_c=\{Tx_c^1, Tx_c^2, ..., Tx_c^n\}$. 
 Before transmitting the transactions to the blockchain, the parameters of these transactions need to be generated, as discussed in Sec. \ref{Indistinguishable Transaction Feature}.


\begin{algorithm} 
\footnotesize
\setstretch{1.35}
\renewcommand{\algorithmicrequire}{\textbf{Input:}}
\renewcommand{\algorithmicensure}{\textbf{Output:}}
\caption{Embed the covert message into $Tx_c$} 
\label{alg:message embedding} 
	\begin{algorithmic}[1]
		\REQUIRE $Esk_{AB}, Message, index_{last}$
		\ENSURE $Tx_c$
	
		\STATE $init: msg \gets \{\}, Tx_c \gets \{\}, index \gets index_{last}$
		\STATE $msg \gets MsgProcessing(Message)$
		\STATE $msg = \{msg_1,...,msg_n\}$
  
		\STATE $i \gets 1$
		\WHILE{$i \leq n$}

        \STATE $sk_c^{index} \gets GenKeyByIndex(Esk_{AB}, index)$
		\STATE $Tx_c^{index} \gets GenTxBySubliminal(sk_c^{index}, msg_i)$
		\STATE add $Tx_c^{index}$ to $Tx_c$
		\STATE $i \gets i+1$
        \STATE $index \gets index+1$
		
        \ENDWHILE
		
		\RETURN $Tx_c$
	\end{algorithmic} 
\end{algorithm}

\noindent
$\bullet$
\textbf{Covert Message Extraction.}
 Algorithm \ref{alg:message extraction} outlines the process by which Bob retrieves covert transactions and extracts the embedded message. 
 Firstly, Bob generates a private key $sk_c^{index}$ using $GenKeyByIndex$. 
 Subsequently, Bob utilizes $GenAddrFromSk$ and $GetTxFromAddr$ to generate the address corresponding to $sk_c^{index}$ and determine if a transaction is associated with this address. 
 Using $GetMsgFromSign$, Bob employs $sk_c^{index}$ to extract the message segment, $msg_i$, from the obtained transaction $Tx_c^{index}$. 
 Finally, Bob reconstructs the complete message from the collected segments.

\begin{algorithm} 
\footnotesize
\setstretch{1.35}
\renewcommand{\algorithmicrequire}{\textbf{Input:}}
\renewcommand{\algorithmicensure}{\textbf{Output:}}
\caption{Extract the covert message from $Tx_c$} 
\label{alg:message extraction} 
	\begin{algorithmic}[1]
		\REQUIRE $Esk_{AB}, index_{last}$
		\ENSURE $Message$
		
		\STATE $init: msg \gets \{\}, index \gets index_{last}$
        \STATE $sk_c^{index} \gets GenKeyByIndex(Esk_{AB}, index)$
		\STATE $addr_c^{index} \gets GenAddrFromSk(sk_c^{index})$ 
		\STATE $Tx_c^{index} \gets GetTxFromAddr(addr_c^{index})$
		\STATE $i \gets 1$
  
		\WHILE{$Tx_c^{index}\ is\ not\ NULL$}
  
    		\STATE $msg_i \gets GetMsgFromSign(Tx_c^{index}, sk_c^{index})$
    		\STATE add $msg_i$ to $msg$
            \STATE $i \gets i+1$
            \STATE $index \gets index+1$
    		
            \STATE $sk_c^{index} \gets GenKeyByIndex(Esk_{AB}, index)$
    		\STATE $addr_c^{index} \gets GenAddrFromSk(sk_c^{index})$ 
    		\STATE $Tx_c^{index} \gets GetTxFromAddr(addr_c^{index})$
  
		\ENDWHILE
	    
	    \STATE $Message \gets msg$
        \RETURN $Message$	    
	\end{algorithmic} 
\end{algorithm}
\section{Evaluation}
\label{Evaluation}
 To assess the concealment and transmission efficiency of \texttt{ABC-Channel}, we address the following research questions.
 
 \begin{mybullet}
     \item 
     \textbf{RQ1: }Can attackers detect covert transactions in black-box?
     
     \item 
     \textbf{RQ2: }Can attackers detect covert transactions in white-box?
     
     \item 
     \textbf{RQ3: }How covert is \texttt{ABC-Channel}? 
     
     \item 
     \textbf{RQ4: }What are the transmission efficiency and cost of the \texttt{ABC-Channel}? 
     
 \end{mybullet}

\subsection{System Implementation}
\label{System Implementation}
 To evaluate the viability of \texttt{ABC-Channel}, we test it on the Bitcoin testnet  (i.e., Testnet), a blockchain designed for developers to experiment and test their projects.
 Unlike the Bitcoin main network (i.e., Mainnet), coins on the Testnet hold no real-world value, ensuring that our experiments do not impact the Mainnet while minimizing experimental costs.
 Table \ref{tab:testing on bitcoin testnet} outlines the construction of four carrier transactions, comprehensively demonstrating the \texttt{ABC-Channel}'s workflow within the blockchain environment.
 
 In the key transaction $Tx_k$, $addr_{Alice}$ is the sending address, and $addr_{Charlie}$ and $addr_{Dave}$ are the receiving addresses, respectively.
 Bob uses $sk_{Bob}$ to get $sk_{Alice}$ from $Tx_k$.
 As described in Sec. \ref{Contactless Channel Negotiation}, Alice and Bob currently generate $Esk_{AB}$.
 Then, the addresses derived by $Esk_{AB}$ (e.g., $addr_c^{67890H}$) are used as the sending addresses of $Tx_c$.
 In $addr_c^{67890H}$, the number means index, and $H$ means hardened derivation.
 This experiment includes two instances of $Tx_c$: the first involving a single input capable of conveying a 32-byte message and the second incorporating two inputs, thus capable of carrying a 64-byte message.

 \begin{table*}[htp]
 \centering
 \setstretch{1.15}
 \small
 \vspace{-7pt}
 \caption{Testing on bitcoin testnet}
 \label{tab:testing on bitcoin testnet}
 \vspace{-5pt}
 \begin{tabular}{llll}
 \hline
 \multirow{7}{*}{$Tx_k$}
  & Txid & ($Tx_k^1$) & 462f093130e615af1373fe82f7cc951e17837c171f4aa9eaa762f50437b20d55 \\
  & Sender & ($addr_{Alice}$) & 	mjmuzfmtguwx3QrGpTmucfkyj9oEQ6kBkd \\
  & Recevier & ($addr_{Charlie}$) & moDWFJ2Cm7PvhCUsD7mAeSAWG4pVBPc1oj \\
 \cline{2-4}
 
  & Txid & ($Tx_k^2$) & a0df4f2f8df73046eac36d3bda8398e6a24b4b8b0b8aa4549b8fdd2a5efacbc5 \\
  & Sender & ($addr_{Alice}$) & mjmuzfmtguwx3QrGpTmucfkyj9oEQ6kBkd \\
  & Recevier & ($addr_{Dave}$) & n1XCWXKZepTkB1JY7vV8KpxGqGTLmcBpHQ \\
  & Info & ($sk_{Alice}$) & cP4tQrMiduNh3tLxFGuMW599YCbkozQ6d1cgAenfYUi8muvsjyZP \\
 \hline
 
  \multirow{10}{*}{$Tx_c$}
  & Txid & ($Tx_c^1$) & 9a359ee38e5f53ab09340f67f7e49b7d50274342c13cd1507965e6c90c0d416e \\
  & Sender & ($addr_c^{67890H}$) & 	mvGSrebnjaMFgNRsiF1YVYCYP7S4pCVNDX \\
  & Recevier & (-) & mnZJJnFc9BvM4XhkaryxkqoSaaLfHwWJCF \\
  & Info & (32 Bytes) & aaaabaaacaaadaaaaaaabaaacaaadaaa \\
 \cline{2-4}
  & Txid & ($Tx_c^2$) & f47449708a94fb740d811c75399d04deae1db6c684313d4d2b27c3bc4309f8ba \\
  & Sender1 & ($addr_c^{55162H}$) & 	mzew8QF2gxGDcdw73D4WaBt6cLqGo9YwzW \\
  & Sender2 & ($addr_c^{37335H}$) & 	mi4YLWjZwAgfNu3w2s9Wfp5RUi9nk4ZsBB \\
  & Recevier1 & (-) & myp6EfxHKgvWcgqxFKfk7KNHdnCG5widZg \\
  & Recevier2 & (-) & mnZJJnFc9BvM4XhkaryxkqoSaaLfHwWJCF \\
  & Info & (64 Bytes) & aaaabaaacaaadaaaaaaabaaacaaadaaaeaaafaaagaaahaaaeaaafaaagaaahaaa \\
 \hline

 \end{tabular}
 \vspace{-5pt}
 \end{table*}

\subsection{Data Collection}
\label{Data Collection}
 To answer \textbf{RQ1}, \textbf{RQ2}, and \textbf{RQ3}, we collected four datasets from Mainnet for evaluating \texttt{ABC-Channel}'s performance.

 \noindent
 \textbf{BTD129 and BTD30 datasets: }
 Blockchair provides historical bitcoin transaction data \cite{blockchair}.
 In June 2022, a dataset comprising approximately 7.5 million transactions was collected, excluding coinbase transactions. 
 Coinbase transactions generated by the block itself rather than by users are thus excluded from our statistical analysis.
 The maximum values for the parameters inputCnt and outputCnt were 1,474 and 3,426, respectively. 
 This dataset was further divided into two subsets: BTD129, consisting of transactions from June 1 to June 29, and BTD30, consisting of transactions from June 30.
 Further dataset analysis revealed that approximately 6.82 million transactions, accounting for about 90.75\%. 
 Consequently, to diminish the influence of outlier data, subsequent experiments focused solely on transactions with inputCnt and outputCnt values of 5 or less.
 

 \noindent
 \textbf{BTD8808 and BAD8808 datasets: }
 Michalski \textit{et al.} \cite{DVN/KEWU0N_2020} published a Bitcoin address dataset comprising 8808 addresses along with their corresponding types. 
 From this dataset, we collected transactions related to these addresses from the Mainnet, up to 500 transactions per address. 
 We use Python to write scripts and get them through API provided by third parties \cite{Blockchain_ledger_API}.
 This resulted in a dataset of approximately 827,000 transactions, referred to as BTD8808. 
 By leveraging the transaction features within BTD8808, we generated relevant address features. 
 As a result, we obtained a dataset containing 8808 address features, which we denote as BAD8808.
 


 We try to use different methods to generate covert transactions.
 According to whether indistinguishable transaction features and untraceable communication identities are used, these methods can be divided into the following four types.

 \noindent
 \textbf{Type S: Sending directly.} 
 Do not use indistinguishable transaction features and untraceable communication identities.
 This type generates a random number of transactions for each address and generates their parameters within a specific range.

 \noindent
 \textbf{Type D: Dynamic address.}
 Only untraceable communication identity is used.
 The difference from  type S is that covert transactions generated by this type  have different addresses.

 \noindent
 \textbf{Type I: Impersonating identity.}
 Only an indistinguishable transaction feature is used.
{ We select transactions whose address type is ``exchange'' from BTD8808 as training data,   and then generate covert transactions of the same type.}

 \noindent
 \textbf{Type  A: \texttt{ABC-Channel}.}
 Simultaneous use of indistinguishable transaction features and untraceable communication identity.
 We use BTD129 as training data for indistinguishable transaction features.
 To avoid the impact of extreme data on training, we removed the data less than 1\% and greater than 99\% according to the percentile of inputsAmount.
 The dataset is divided into three parts based on the percentile of inputsAmount: 1\%-20\%, 20\%-80\%, and 80\%-99\%. 
 The GAN network is trained on each of these parts separately.


\subsection{Black-box Testing}
\label{Black-box Testing}
 Without knowledge of \texttt{ABC-Channel}'s workflow, attackers can only detect covert transactions through black-box testing.

\noindent
$\bullet$
\textbf{Experimental Setup.}
 This experiment utilizes the BTD30 and BAD8808 datasets as the foundation. 
 To enable a thorough evaluation, we amalgamate real transactions with covert transactions to form mixed datasets. 
 In each mixture, covert transactions constitute approximately 50\% of the total.
 The specific mixing method depends on the threat posed by attackers, which can be categorized as detecting the addresses used by covert transactions (threat 1) or individual covert transactions (threat 2).

 Due to their lack of untraceable communication identities, both type S and type I are vulnerable to threat 1, as attackers can extract address features. 
 Once the addresses used in these methods are identified, discovering covert transactions becomes feasible for attackers. 
 To simulate this scenario, we extract the address features of type S and type I and incorporate them into the BAD8808 dataset. 
 Type I is specifically designed to emulate exchange-type transactions. 
 Hence, when creating the mixed dataset, only exchange-type transactions from the BAD8808 dataset are retained.
 Type D and type A primarily face threat 2. 
 Therefore, we mix them separately into the BTD30 dataset.
 

 This subsection aims to employ k-means to identify covert transactions within these mixed datasets.

\noindent
$\bullet$
\textbf{Result.}
 Table \ref{tab:ABC-Channel ablation experiment} displays the experimental results, where we perform clustering using k-means and evaluate the results using metrics such as Adjusted Rand Index (ARI) and Normalized Mutual Information (NMI).
 In our experiment, we set the number of clusters (k) in k-means to 2 to distinguish between covert and real transactions.  
 The findings indicate that attackers are unable to discern covert transactions irrespective of the technique used to generate them.

\answer{1}{
 In black-box testing, k-means clustering of covert transactions from \texttt{ABC-Channel} yields an ARI of 0.0 and an NMI of 0.002, indicating that attackers cannot detect covert transactions.
}

\begin{table*}[htp]
\vspace{-7pt}
\caption{Ablation experiment (M1: indistinguishable transaction feature \& M2: untraceable communication identity)}
\vspace{-5pt}
\label{tab:ABC-Channel ablation experiment}
\centering
\setstretch{1.2}
\scriptsize
\resizebox{\textwidth}{!}{
\begin{tabular}{lcccccccccccc}
\hline
\multirow{2}{*}{Type} & \multirow{2}{*}{M1} & \multirow{2}{*}{M2} & \multirow{2}{*}{Real Dataset} & \multirow{2}{*}{Threat} & \multicolumn{2}{c}{K-means} & \multicolumn{3}{c}{Random Forest} & \multicolumn{3}{c}{GBDT} \\ \cline{6-13} 

 & & & & & ARI & NMI & Precision & Recall & F1-score & Precision & Recall & F1-score \\ \hline
 Type S: Sending directly & w/o & w/o & BAD8808 & threat 1  & 0.0 & 0.021 & 1.0 & 1.0 & 1.0 & 1.0 & 1.0 & 1.0 \\ 
 Type D: Dynamic address & w/o & w/ & BTD30 & threat 2 & 0.0 & 0.002 & 0.933 & 0.943 & 0.938 & 0.919 & 0.944 & 0.931 \\ 
 Type I: Impersonating identity & w/ & w/o & BAD8808 & threat 1 & 0.0 & 0.035 & 1.0 & 1.0 & 1.0 &1.0 & 1.0 & 1.0 \\ \hline
 Type A: \texttt{ABC-Channel} & w/ & w/ & BTD30 & threat 2 & 0.0 & 0.002 & 0.606 & 0.608 & 0.607 & 0.578 & 0.593 & 0.586 \\ \hline
\end{tabular}
}
 \vspace{-5pt}
\end{table*}

\subsection{White-box Testing}
\label{White-box Testing}
In this subsection, we conduct a white-box evaluation of \texttt{ABC-Channel}.

\noindent
$\bullet$
\textbf{Experimental Setup.}
 The setup for this evaluation mirrors that described in Sec. \ref{Black-box Testing},  with the primary distinction being the assumption that attackers have gained access to \texttt{ABC-Channel}.  
 This access enables attackers to create labeled data for training and testing, leading to the development of classifiers, such as random forest and gradient boosting decision tree (GBDT), capable of identifying covert transactions.
 

\noindent
$\bullet$
\textbf{Result.}
 Table \ref{tab:ABC-Channel ablation experiment} presents the results of our experiments, where we evaluate \texttt{ABC-Channel} using precision, recall, and F1-score as metrics.
 We observe that the covert transactions generated by type S and type I are all detected.
 Despite our efforts to obscure each covert transaction by type I, the effectiveness of this approach was not significant.
 This outcome can be attributed to the inherent difficulty in disguising address features, which remain identifiable and can be used by attackers to detect covert transactions. 

 We use transaction frequency as an example to illustrate the challenges posed by address features in the blockchain covert channel.
 Transaction frequency is calculated as the number of transactions corresponding to an address divided by the range of block heights.
 Through statistical analysis of the BAD8808 dataset, we observe that different types of addresses generally exhibit low transaction frequencies.
 In the blockchain covert channel, this low transaction frequency directly affects communication efficiency.
 Furthermore, several other address features are challenging to simulate, such as the presence of coinbase transactions and the number of senders and receivers in each transaction.
 Untraceable communication identity in type A can avoid these issues and ensure effective covert communication.

 The type D performs slightly better than type S and type I.
 However, type D lacks indistinguishable transaction features, resulting in covert transactions with fixed parameters that deviate significantly from real transactions.
 Type A combines untraceable communication identity and indistinguishable transaction features, achieving the highest level of concealment for high-throughput message transmission.
 By incorporating these modules, type A outperforms the mentioned methods in overall concealment effectiveness.


\answer{2}{
 In white-box testing, \texttt{ABC-Channel} achieved precision, recall, and F1-score of 0.606, 0.608, and 0.607 with random forest, and 0.578, 0.593, and 0.586 with GBDT. 
 The experimental results demonstrate that attackers cannot accurately detect covert transactions.
}


 To answer \textbf{RQ3},  we conducted comparative experiments across various datasets, as depicted in Table \ref{tab:comparative experiment}.
 We used random guesses as a benchmark and random forest as the classifier.
 The closer the classification result is to random guess, the stronger the concealment.
 We keep the ratio of positive and negative samples around 1:1.
 To ensure a fair comparison, we adjusted the recall to 0.5 and maintained a balanced ratio of positive and negative samples.
 The precision of the second experiment reached 0.626, which was 0.126 higher than the random guess. 
 This improvement can be attributed to \texttt{ABC-Channel} using BTD129 as a training set.
 In the third experiment, we differentiated between two datasets of real transactions, and the results showed distinguishable transaction patterns across different periods, supporting our findings.

\begin{table}[htp]
\vspace{-7pt}
\caption{Comparative experiment}
\vspace{-5pt}
\label{tab:comparative experiment}
\centering
\resizebox{\linewidth}{!}{
\begin{tabular}{cccccc}
\hline
\multirow{2}{*}{Method} & \multirow{2}{*}{Negative} & \multirow{2}{*}{Positive} & \multicolumn{3}{c}{Metric} \\ \cline{4-6} 
 & & & \multicolumn{1}{c}{Precision} & \multicolumn{1}{c}{Recall} & F1-score \\ \hline
 Random Guess & - & - & \multicolumn{1}{c}{0.5}  & \multicolumn{1}{c}{0.5}  & 0.5 \\ 
 Random Forest & BTD30 & \texttt{ABC-Channel} & \multicolumn{1}{c}{0.626}  & \multicolumn{1}{c}{0.501}  & 0.557  \\
 Random Forest & BTD30 & BTD129 & \multicolumn{1}{c}{0.598}  & \multicolumn{1}{c}{0.491}  & 0.539  \\ \hline
 Random Forest & BTD129 & \texttt{ABC-Channel} & \multicolumn{1}{c}{0.549}  & \multicolumn{1}{c}{0.491}  & 0.518  \\ \hline
\end{tabular}
}
\vspace{-10pt}
\end{table}

 The results of the fourth experiment indicate that the covert transactions generated by \texttt{ABC-Channel} are challenging to differentiate from the training data.
 It should be noted that our training set is limited to historical transaction data, which may exhibit variations over time.
 Moreover, we cannot anticipate future changes, such as fluctuations in transaction fees or Bitcoin's price.
 Therefore, under these limitations, \texttt{ABC-Channel} has achieved a desirable level of concealment.


\answer{3}{
 When the recall is set to 0.5, and the training data (BTD129) is treated as negative samples, the precision of \texttt{ABC-Channel} is 0.549.
 This precision is close to the 0.5 benchmark of random guess.
 This indicates that \texttt{ABC-Channel} provides adequate concealment.
}

\subsection{Impact of Transaction Parameters on Concealment}

We examine the influence of evaluating transaction parameters on transaction concealment. 
 We investigate methods rooted in Bitcoin, particularly those leveraging subliminal channels \cite{tian2020dlchain, frkat2018chainchannels, tiemann2021act}. 
 Existing research often entails confirmatory experiments without meticulously considering the transaction parameter settings. 
 Therefore, adhering to specific rules, we generate a substantial volume of specialized transactions to simulate high-throughput communication scenarios.

 The number of inputs (inputCnt) and outputs (outputCnt) in covert transactions are generated randomly, with values ranging from 1 to 5. 
 The amounts for inputs (inputsAmount) and transaction fees are classified into five categories based on percentile ranges derived from dataset BTD129, namely p1 (1\%-20\%), p2 (20\%-40\%), p3 (40\%-60\%), p4 (60\%-80\%), and p5 (80\%-99\%). 
 For each covert transaction, the inputsAmount and fee are generated randomly but within the specific percentile range. 
 The amount of outputs (outputsAmount) is determined by subtracting the transaction fee from the inputsAmount. 
 To construct a balanced dataset, BTD30 was used as the control group, achieving a roughly 1:1 ratio between positive and negative samples. 
 The Random Forest algorithm was selected for the experimental analysis, and the F1-score was utilized as the primary metric for evaluation.

 We conducted a total of 25 experiments, as depicted in Table \ref{tab:different transaction parameter intervals}. 
 It is apparent that if users of covert channels adopt a method of generating covert transactions with fixed-range parameters, there is a high probability of exposure for these covert transactions when engaging in high-throughput covert communication. 
 In contrast, \texttt{ABC-Channel} achieved an F1-score of 0.607 in the same test. 
 Experimental results indicate that the parameter generation method proposed in this paper can effectively enhance the concealment of covert transactions.


\begin{table}[htp]
\footnotesize
\caption{Recognition results of covert transactions generated using different transaction parameter intervals}
\vspace{-8pt}
\label{tab:different transaction parameter intervals}
\centering
\begin{tabular}{c|ccccc}
\hline
\diagbox[width=2.3cm, height=0.8cm]{inputsAmount}{fee} & p1    & p2    & p3    & p4    & p5    \\ \hline
p1  & 0.987 & 0.983 & 0.987 & 0.990 & 0.991 \\
p2  & 0.985 & 0.982 & 0.983 & 0.986 & 0.988 \\
p3  & 0.985 & 0.980 & 0.982 & 0.981 & 0.984 \\
p4  & 0.987 & 0.981 & 0.982 & 0.981 & 0.983 \\
p5  & 0.997 & 0.995 & 0.994 & 0.993 & 0.994 \\ \hline
\end{tabular}
\vspace{-10pt}
\end{table}

\subsection{Transmission Efficiency Analysis}
\label{Transmission Efficiency Analysis}

 In blockchain-based covert channels, the amount of information that can be transmitted in a transaction is considered the measure of transmission efficiency. 
 To ensure comparability, we focus on the scenario where a transaction has a single input.
 In this scenario, implicit embedding allows a transaction to carry a message of up to 256 bits.
 We refer to methods that do not directly use public modifiable fields as implicit embedding.
 In \texttt{ABC-Channel}, a transaction with one input can carry a 256-bit message, achieving a SOTA-level transmission efficiency.
 We compare the transmission efficiency of \texttt{ABC-Channel} with other methods in Table \ref{tab:transmission efficiency comparison}.

 While the method proposed by Gao \textit{et al.} \cite{gao2020achieving} allows carrying a 640-bit message in a transaction, it lacks expandability.
 We introduce expandable capacity as a metric to assess whether increasing the number of inputs can improve the transmission efficiency of a single transaction.
 Methods like \texttt{ABC-Channel} can enhance transmission efficiency in this way.
 The transmission efficiency of these methods can be influenced by the blockchain platform used. 
 For example, in Monero, an input with multiple signatures can enhance transmission efficiency \cite{lan2020using}.


 
\begin{table}[htp]
\footnotesize
\vspace{-7pt}
\caption{Transmission efficiency comparison}
\vspace{-5pt}
\label{tab:transmission efficiency comparison}
\centering
\resizebox{\linewidth}{!}{
\begin{tabular}{p{2.2cm}<{}p{1.18cm}<{\centering}p{2.28cm}<{\centering}p{1.25cm}<{\centering}p{1.33cm}<{\centering}}
\hline
    Literature & \makecell[c]{Capability\\(bits)} & \makecell[c]{Message\\carrier} & \makecell[c]{Implicit\\embedding} & \makecell[c]{Expandable\\capacity}\\ \hline
    
    Gao et al. \cite{gao2020achieving} & 640 & \makecell[c]{OP\_RETURN} & $\times$ & $\times$ \\ 

    MRCC \cite{guo2021mrcc} & 11 & \makecell[c]{LSB of public keys} & $\times$ & \checkmark \\ 
        
    MSCCS \cite{liu2022msccs} & 39 & \makecell[c]{Transaction amount} & \checkmark & \checkmark \\ 
    
    DLchain \cite{tian2020dlchain} & 128 & Nonce of signature & \checkmark & \checkmark \\ 
    
    Tiemann et al. \cite{tiemann2021act} & 256 & \makecell[c]{Nonce of signature} & \checkmark & \checkmark \\ \hline
    
    \texttt{ABC-Channel} & 256 & \makecell[c]{Nonce of signature} & \checkmark & \checkmark \\ \hline
    
\end{tabular}
}
\vspace{-6pt}
\end{table}
 In practical scenarios, the issue of stealth must be taken into account. 
 Therefore, we have analyzed the distribution of transactions with different input counts in BTD129, along with the corresponding average transaction fees, as shown in Table \ref{tab:proportion of input}. 
 Calculating the expected values from these figures can be considered a measure of the transmission efficiency of the \texttt{ABC-Channel} in real-world contexts.


\begin{table}[htp]
\footnotesize
\caption{Transaction proportion and average fee with different input numbers}
\vspace{-5pt}
\label{tab:proportion of input}
\centering
\begin{tabular}{p{1.8cm}<{}ccccc}
\hline
Input number & 1        & 2        & 3        & 4        & 5        \\ \hline
Proportion   & 0.777    & 0.140    & 0.047    & 0.022    & 0.014    \\ \hline
Average fee  & 3144.32 & 4373.07 & 5785.43 & 7241.43 & 7393.79 \\ \hline
\end{tabular}
\vspace{-10pt}
\end{table}

 We can calculate that, on average, each transaction in BTD129 has 1.355 inputs with a fee of 3589 Satoshi. 
 Therefore, when employing the \texttt{ABC-Channel} for large-scale covert communication, the transmission efficiency is approximately 347 bit/$Tx_c$, and the cost is about 3589 Satoshi/$Tx_c$.

\answer{4}{
 In \texttt{ABC-Channel}, each covert transaction with $n$ inputs can carry a $256\times n$-bit message, a SOTA level compared to other methods using implicit embedding in Bitcoin.
 In practical scenarios, the transmission efficiency is 347 bits/$Tx_c$, and the cost is 3589 Satoshi/$Tx_c$.
}

\section{Discussion}
\label{Discussion}
 This section discusses potential challenges that \texttt{ABC-Channel} may encounter in practical scenarios. 

\noindent
$\bullet$
\textbf{Off-chain Information Leakage.}
 Off-chain information, such as the bank card used for cryptocurrency cashing or payment records for purchasing cryptocurrencies, may provide clues for de-anonymization. 
 Since the inception of Bitcoin, research on the de-anonymization of Bitcoin addresses has been continuously evolving \cite{li2020identifying, koshy2014analysis, wallace2020can, biryukov2014deanonymisation}. 
 Meiklejohn \textit{et al.} \cite{meiklejohn2013fistful} proposed a heuristic approach to identify user categories by collecting marked addresses and constructing transaction graphs. 
 This is not an issue unique to Bitcoin. 
 Even privacy-focused cryptocurrencies such as Monero and Zcash face threats of de-anonymization \cite{biryukov2019deanonymization}. 
 The primary reason is that, regardless of the type of cryptocurrency, users inevitably leak information during usage. 
 Therefore, all blockchain-based covert channels may face threats from off-chain information leakage in real-world scenarios.


\noindent
$\bullet$
\textbf{Covert Transaction Withdrawal.}
 Researchers have proposed using double spending to avoid including covert transactions in the blockchain. 
 They argue that this approach enhances the concealment of covert channels since attackers cannot detect these covert transactions within historical transaction data \cite{yin2020coinbot, huang2021covert}. 
 This method proves effective if attackers attempt to identify transactions that have already been packaged into blocks. 
 However, this method becomes ineffective if attackers monitor all transactions sent to the blockchain. 
 Most regular users refrain from engaging in double spending, and if Alice were to employ this technique, it would attract attackers' attention. 
 Moreover, whether or not covert transactions are included in the blockchain, they are still broadcast. 
 Consequently, double spending draws the attention of attackers and fails to prevent them from identifying covert transactions.


\noindent
$\bullet$
\textbf{Simulate More Transaction Parameters.}
\label{More transaction Fields}
 In Sec. \ref{Indistinguishable Transaction Feature}, we concentrated on five parameters employed in Bitcoin transactions. 
 These parameters are pivotal as they necessitate user specification for each transaction. 
 Besides these, there are parameters with specialized functions, like locktime and sequence, which, in contrast to the primary parameters, often have default settings and seldom need adjustments. 
 For example, the locktime parameter enables users to determine the precise timing for a transaction's inclusion in the blockchain. 
 Consequently, our research prioritizes the generation of covert transactions using these predominant parameters over a comprehensive examination of all available parameters.
 

\section{Related Work}
\label{Related Work}

 Existing research on blockchain-based covert channels primarily focuses on message embedding techniques.
 One common approach is to utilize modifiable fields within the blockchain, such as OP\_RETURN, to carry the covert messages. 
 For instance, Gao \textit{et al.} \cite{gao2020achieving} proposed a method that employs kleptography for transmitting and decrypting private keys through modifiable fields in blockchains such as Bitcoin and Ethereum.
 Partala \textit{et al.} \cite{partala2018provably} introduced BLOCCE, a technique that utilizes the least significant bit (LSB) of addresses to carry a 1-bit message. 
 Also, there are many attempts to exploit blockchains for controlling botnets \cite{ali2018zombiecoin, franzoni2020leveraging, yin2020coinbot}. 
 Although this kind of method is straightforward, it lacks concealment.

 To enhance the concealment of blockchain-based covert channels, many methods have employed implicit embedding. 
 Tiemann \textit{et al.} \cite{tiemann2021act} introduced a novel approach for disclosing nonces in signatures.
 Tian \textit{et al.} \cite{tian2020dlchain} proposed DLchain, which utilizes digital signatures as message carriers and dynamically labels them using the OP\_RETURN field. 
 In DLchain, the subliminal channel enables two digital signatures to carry a 32-byte message. 
 Additionally, some approaches leverage transaction relationships for message transmission. 
 Cao \textit{et al.} \cite{cao2020chain} proposed utilizing the linkage between public keys, while Luo \textit{et al.} \cite{luo2021novel} suggested leveraging the interaction relationship between addresses. 
 Implicit embedding techniques offer superior concealment compared to explicit embedding approaches.
 However, these methods do not solve the problems caused by high-throughput covert communication.

 Using fixed addresses in covert channel communication poses an increased risk of exposure. 
 Deanonymization research in the blockchain domain has already emerged \cite{biryukov2019deanonymization, gaihre2018bitcoin, biryukov2019transaction}.
 Xiang \textit{et al.}  \cite{xiang2022babd} have employed machine learning techniques to classify Bitcoin addresses based on their behavior. 
 Tiemann \textit{et al.} \cite{tiemann2021act} proposed a method where Alice and Bob agree on a receiving address, but this still encounters the challenge of fixed addresses. 
 Alternatively, Alice and Bob can choose not to agree on the receiving address, which would incur additional costs. 
 Additionally, off-chain communication carries its own risks. 
 Encryption can protect negotiation information from leakage but does not guarantee the concealment of communication identities. 
 Even in insecure network environments, using the Tor network does not guarantee anonymity \cite{nasr2018deepcorr}.

 Establishing a covert channel based on blockchain characteristics is a research focus.
 Monero uses ring signatures and ring confidential transactions to protect the identities and transaction amounts of the parties involved.
 MRCC and MSCCS utilize the input address and transaction amount as message carriers \cite{guo2021mrcc, liu2022msccs}. 
 In addition, Ethereum, Zcash, and Bitcoin Lightning Network are also used to build covert channels \cite{liu2020whispers, baden2019whispering, gimenez2021zephyrus, biryukov2019privacy, kurt2020lnbot}.
 However, these methods do not apply to a wide range of blockchain platforms.

 In a departure from preceding studies, our \texttt{ABC-Channel} framework is tailor-made to tackle the challenge of facilitating high-throughput covert communications. 
 Moreover, it investigates strategies for establishing covert channels that circumvent the need for direct interactions between the sender and recipient. 
 Crucially, the methodologies proposed within the \texttt{ABC-Channel} are designed to be independent of any specific blockchain infrastructure, underscoring their versatility and broad applicability.
 

\section{Conclusion}
\label{Conclusion}
 
We proposed \texttt{ABC-Channel}, a blockchain-based covert channel that fulfills three essential requirements: i) contactless channel negotiation, ii) indistinguishable transaction features, and iii) untraceable communication identity.
 Besides being highly secure, \texttt{ABC-Channel} also effectively achieved high-throughput covert communication. 
 We conducted extensive experiments and found that when the recall is 0.491, the classifier can distinguish transactions generated by \texttt{ABC-Channel} with an F1-score of 0.518, which is close to 0.5 (representing a random guess).
 This suggests that attackers cannot detect covert transactions accurately.
 In \texttt{ABC-Channel}, each covert transaction with $n$ inputs can carry a $256\times n$-bit message, a SOTA level compared to other methods using implicit embedding in Bitcoin. 
 Moreover, \texttt{ABC-Channel} is not dependent on specific blockchain platforms, making it applicable to a wide range of blockchain platforms.


\bibliographystyle{IEEEtran}
\bibliography{ref}

\begin{IEEEbiography}[{ \includegraphics[width=1in, height=1.25in, clip, keepaspectratio]{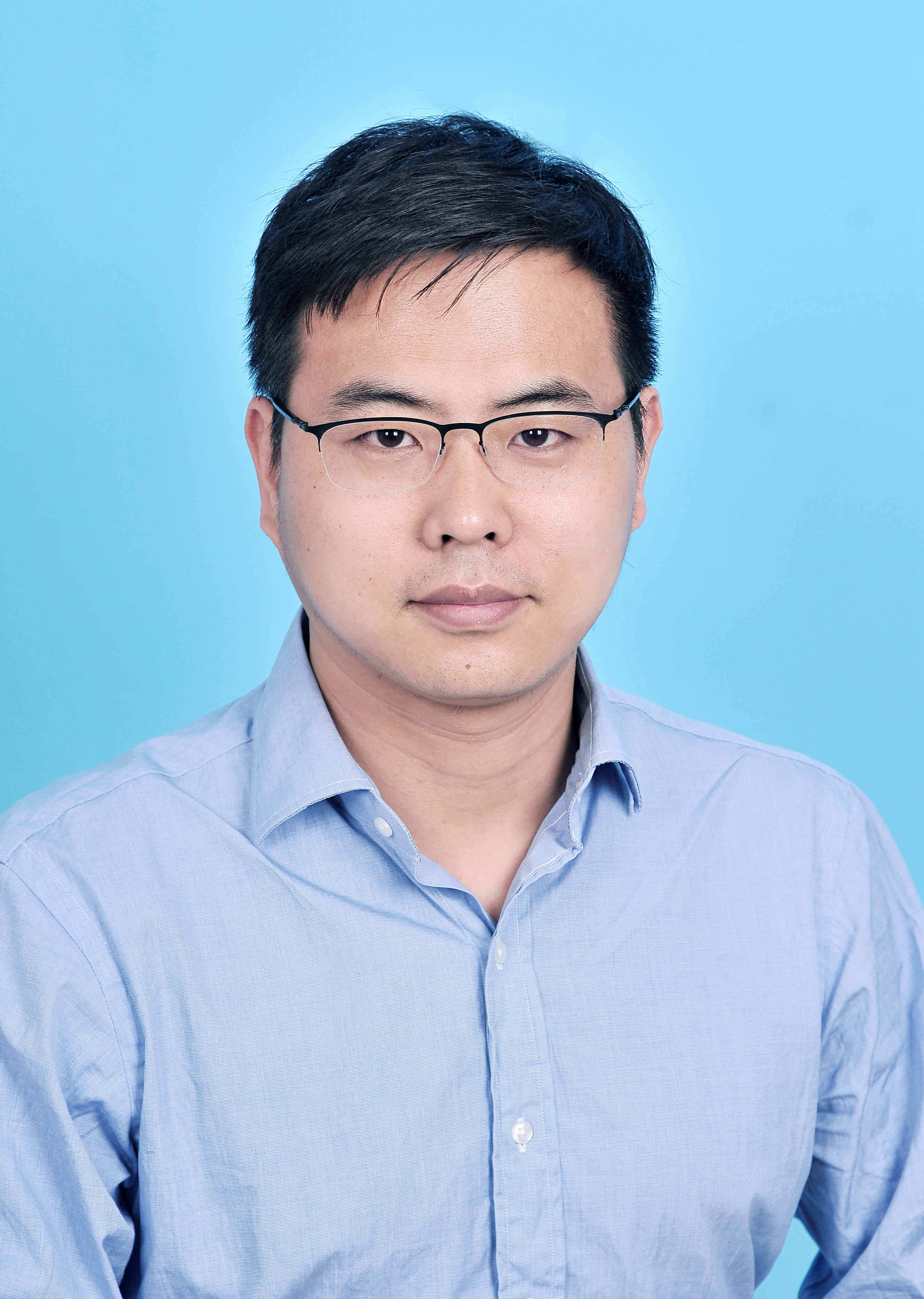}}]{Xiaobo Ma} is a professor with MOE Key Lab for Intelligent Networks and Network Security, Faculty of Electronic and Information Engineering, Xi’an Jiaotong University, Xi’an, China. He received a Ph.D. in Control Science and Engineering from Xi’an Jiaotong University in 2014, and was a Post-Doctoral Research Fellow with The Hong Kong Polytechnic University in 2015. He is a Tang Scholar, and a member of IEEE. His research interests include Internet measurement and cyber security.
\end{IEEEbiography}

\vspace{-1.2cm}

\begin{IEEEbiography}[{ \includegraphics[width=1in, height=1.25in, clip, keepaspectratio]{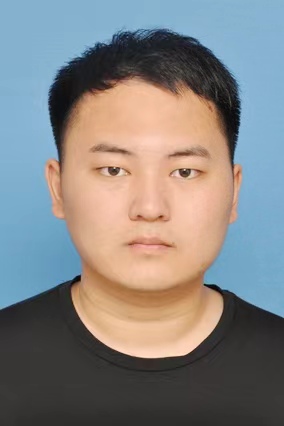}}]{Pengyu Pan} 
received the bachelor’s degree in computer science and technology in 2020. He is currently pursuing the Ph.D. degree with the MOE Key Lab for Intelligent Networks and Network Security, Faculty of Electronic and Information Engineering, Xi’an Jiaotong University, Xi’an, China. His current research interests include blockchain and cyber security.
\end{IEEEbiography}

\vspace{-1.2cm}

\begin{IEEEbiography}[{\includegraphics[width=1in, height=1.25in, clip, keepaspectratio]{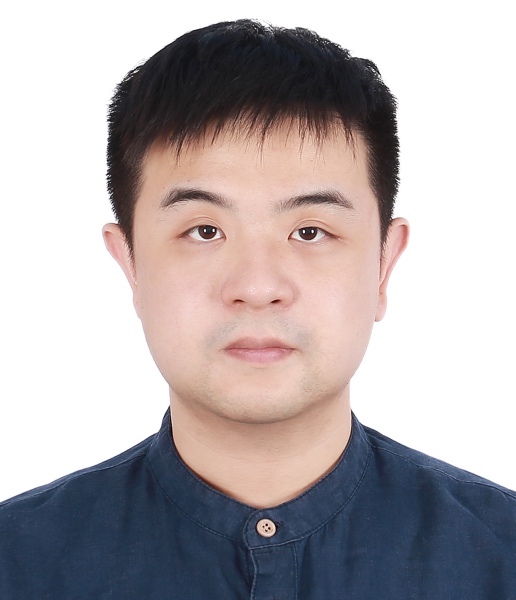}}]{Jianfeng Li} received his Ph.D. degree in Control Science and Engineering from Xi’an Jiaotong University, China, in Mar. 2018. He is an Assistant Professor with the MOE Key Laboratory for Intelligent Networks and Network Security, Faculty of Electronic and Information Engineering, Xi’an Jiaotong University. He worked as a postdoctoral fellow at the Hong Kong Polytechnic University from Sep. 2019 to Jun. 2022. His research interests have centered on traffic analysis, privacy of mobile platform, network monitoring, AI security, and large-scale cyber security. He has published a number of research papers in top conferences and journals, such as S\&P, CCS, USENIX Security, NDSS, INFOCOM, TON, and TIFS.
\end{IEEEbiography}

\vspace{-1.2cm}

\begin{IEEEbiography}[{\includegraphics[width=1in, height=1.25in, clip, keepaspectratio]{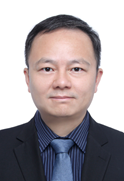}}]{Wei Wang}
is a full Professor with school of computer science and technology, Beijing Jiaotong University, China. He received the Ph.D. degree from Xi'an Jiaotong University, in 2006. He was a Post-Doctoral Researcher with the University of Trento, Italy, from 2005 to 2006. He was a Post-Doctoral Researcher with TELECOM Bretagne and with INRIA, France, from 2007 to 2008. He was also a European ERCIM Fellow with the Norwegian University of Science and Technology (NTNU), Norway, and with the Interdisciplinary Centre for Security, Reliability, and Trust (SnT), University of Luxembourg, from 2009 to 2011. His recent research interests lie in data security and privacy-preserving computation. He has authored or co-authored over 100 peer-reviewed articles in various journals and international conferences, including IEEE TDSC, IEEE TIFS, IEEE TSE, ACM CCS, AAAI, Ubicomp, IEEE INFOCOM. He has received the ACM CCS 2023 Distinguished Paper Award. He is an Elsevier ``highly cited Chinese Researchers''. He is an Associate Editor for IEEE Transactions on Dependable and Secure Computing, and an Editorial Board Member of Computers \& Security and of Frontiers of Computer Science. He is a vice chair of ACM SIGSAC China.
\end{IEEEbiography}

\vspace{-1.2cm}

\begin{IEEEbiography}[{\includegraphics[width=1in, height=1.25in, clip, keepaspectratio]{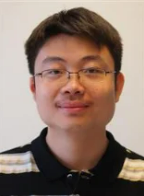}}]{Weizhi Meng}
 is currently an Associate Professor in the Department of Applied Mathematics and Computer Science, Technical University of Denmark (DTU), Denmark. He won the Outstanding Academic Performance Award during his doctoral study, and is a recipient of the Hong Kong Institution of Engineers (HKIE) Outstanding Paper Award for Young Engineers/Researchers in both 2014 and 2017. He also received the IEEE ComSoc Best Young Researcher Award for Europe, Middle East, \& Africa Region (EMEA) in 2020. His primary research interests are cyber security and intelligent technology in security, including intrusion detection, IoT security, biometric authentication, and blockchain. He serves as associate editors / editorial board members for many reputed journals such as IEEE TDSC.
\end{IEEEbiography}

\vspace{-1.2cm}

\begin{IEEEbiography}[{\includegraphics[width=1in, height=1.25in, clip, keepaspectratio]{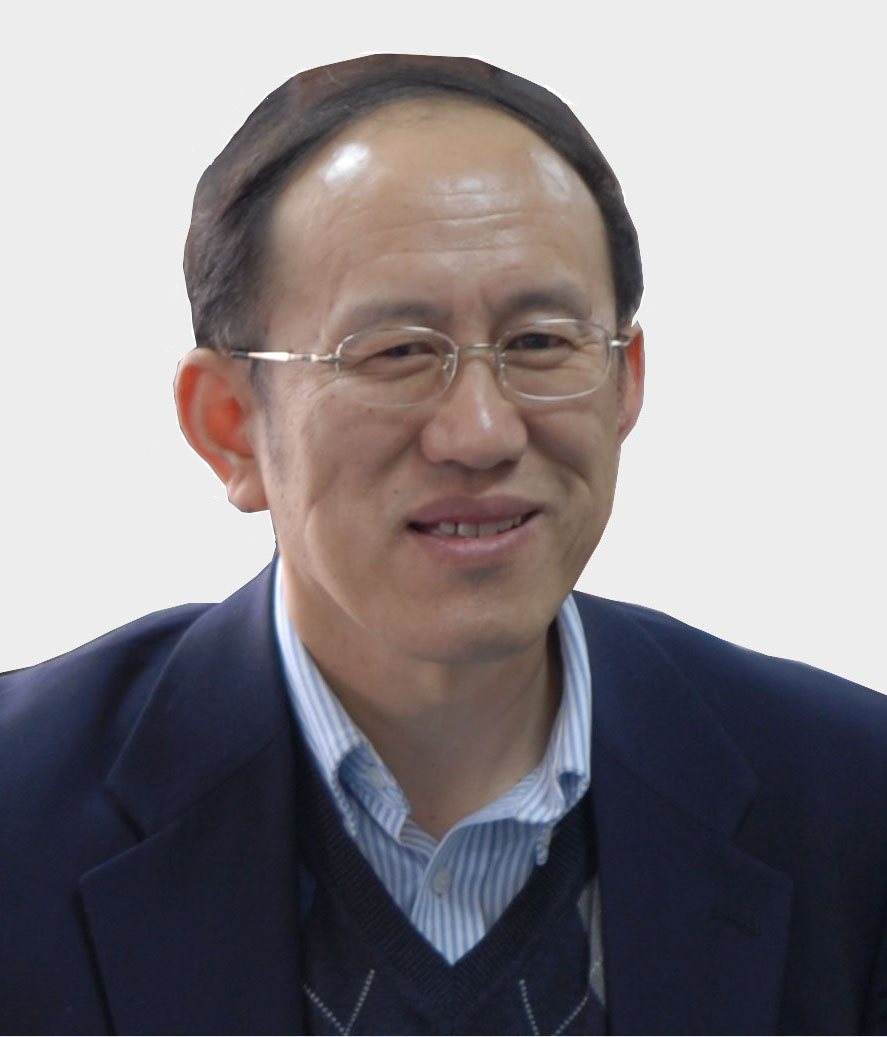}}]{Xiaohong Guan}
	is with 
the MOE Key Lab for Intelligent Networks and Network Security, Faculty of Electronic and Information Engineering,  Xi'an Jiaotong University, Xi'an, China.   
 He is currently 
 the Dean of the Faculty of Electronic and Information Engineering.   
He received the Ph.D. degree in electrical engineering from the University of Connecticut, Storrs, in 1993. Since 1995, he is also with the Department of Automation, Tsinghua National Laboratory for Information Science and Technology, and the Center for Intelligent and Networked Systems,   Tsinghua University. He is an IEEE Fellow and
 an Academician of Chinese Academy of Sciences.
\end{IEEEbiography}
\balance 
\vfill

\end{document}